\documentclass{article}
\usepackage{natbib}
\usepackage{multicol}
\usepackage[figuresright]{rotating}
\usepackage{tikz}
\usepackage{amsmath}
\usetikzlibrary{arrows.meta,
                chains,
                positioning,
                shapes.geometric,shadows,
                }
\usetikzlibrary{automata,positioning}
\usetikzlibrary{shapes,arrows}
\usepackage{tabularx}
\usepackage{float}

\usepackage{arxiv}

\usepackage[utf8]{inputenc} 
\usepackage[T1]{fontenc}    
\usepackage{hyperref}       
\usepackage{url}            
\usepackage{booktabs}       
\usepackage{amsfonts}       
\usepackage{nicefrac}       
\usepackage{microtype}      
\usepackage{lipsum}

\title{How well can screening sensitivity and sojourn time be estimated?}

\author{
  Ayman Hijazy \\
  Department of Probability Theory and Statistics\\
  E\"otv\"os Lor\'and University\\ Budapest, Hungary \\
  Faculty of Informatics\\
  University of Debrecen\\
  Debrecen, Hungary\\
    \texttt{aymanh@cs.elte.hu} 
   \And
  Andr\'as Zempl\'eni \\
Department of Probability Theory and Statistics\\
  E\"otv\"os Lor\'and University\\ Budapest, Hungart \\
  Faculty of Informatics\\
  University of Debrecen\\
  Debrecen, Hungary\\
   \texttt{zempleni@caesar.elte.hu}
\\
}

\begin{document}
\maketitle

\begin{abstract}
Chronic disease progression models are governed by three main parameters: sensitivity, preclinical intensity, and sojourn time. The estimation of these parameters helps in optimizing screening programs and examine the improvement in survival. Multiple approaches exist to estimate those parameters. However, these models are based on strong underlying assumptions. The main aim of this article is to investigate the effect of these assumptions. For this purpose, we developed a simulator to mimic a breast cancer screening program directly observing the exact onset and the sojourn time of the disease. We investigate the effects of assuming the sensitivity to be constant, inter-screening interval and misspecifying the sojourn time. Our results indicate a strong correlation between the estimated parameters, and that the chosen sojourn time-distribution has a strong effect on the accuracy of the estimation. These findings shed a light on the seemingly discrepant results got by different authors using the same data sets but different assumptions. 
\end{abstract}

\twocolumn{}
\section{Introduction}
\label{s:intro}

The natural progression of a disease in the model proposed by \cite{shenzelen} is regarded as a 3 state model: individuals progress from a disease free state $S_f$ to the preclinical state $S_p$, when the disease has become onset but is still asymptomatic, i.e. the person has the disease but it does not show any symptoms. The final state of the disease from our point of view is when it manifests itself through clinical symptoms, thus it is called the clinical state $S_c$.
\begin {center}
\begin{tikzpicture}
\node[state] (s) {$S_f$};
\node[state, right=of s] (r) {$S_p$};
\node[state,right =of r] (c){$S_c$}; 
\draw[every loop]
(s) edge[] node {} (r)
(r) edge node {} (c)
(s) edge[loop above] node {} (s)
(r) edge[loop above] node {} (r);
\end{tikzpicture}
\end{center}

Sojourn time is defined as the amount of time spent in the preclinical state $S_p$, in other words it is the time needed for the disease to show itself by means of clinical symptoms. However, directly observing sojourn time is not feasible as the exact time of onset is unknown.The sojourn time is then estimated through modelling, mostly by assuming it is a random variable with a specified distribution, see e.g. \cite{Wu}, \cite{zelenfe}.

Early detection methods such as screening allows discovering the disease before any symptoms appear. Screening sensitivity, defined as the the probability of detection given that the patient is in $S_p$, is crucial in determining the efficiency of the screening program. 
A case is said to be prevalent if its disease is detected by a screen, and incident or interval if its disease is detected by means of clinical symptoms between two screening rounds. Note that while modelling the process, false positive cases are not of interest, as further medical examination will reveal the absence of the disease.

The parameters of interest in such a process are the preclinical intensity (probability of moving from the disease free state to the preclinical one during $(t,t+dt)$),  the sojourn time and screening sensitivity. The estimation of these parameters is essential to optimize screening intervals and to correct lead time bias. Lead time bias is defined as the apparent increase in survival due to early detection by means of screening. 

The basis for disease progression models was set by \cite {zelenfe} setting the foundations of the theory of periodic screening,  \cite{porok} extended the theoretical background. Later, \cite{shenzelen} introduced two models to estimate the parameters governing disease progression. The first describes stable diseases that are assumed to have incidence and prevalence independent of time or age. The other incorporates time dependence of incidence and prevalence to the model (these cases are called non stable diseases).

These models are built by deriving the probabilities of cases being detected by screening or symptoms, this allows the forming a likelihood function from which parameters can be estimated by classical methods such as nonlinear maximization of the likelihood function, a least squares approach or a Bayesian one.

\cite{Wu} further extended the results by allowing both the transition probability from the disease-free state and the sensitivity to be age dependent. They assume that the sojourn time follows a log-logistic distribution, the preclinical intensity has a log-normal distribution and the sensitivity is age dependent, the age dependence is incorporated by assuming that the sensitivity has a logistic function form. 

\cite{duffychen} proposed a Markov chain model assuming that the rate of transition between the preclinical and the clinical state and the rate of transition between the disease free and the preclinical state are time independent parameters to be estimated. In this model the parameters can be estimated without control data.

In this paper we have constructed a simulator to imitate the results of a breast cancer screening program. Using this tool our main goal was to check and compare the performance of the models under different assumptions. So we simulated using different sojourn time distributions, age dependent sensitivity, and different inter-screening times. The simulator allows recording of the actual sojourn time, something which is not possible in a real life scenario. This made it possible to compare the estimated and the actual values.

The main aim of this paper is to see the effects of choosing an incorrect distribution for the sojourn time, the consequences of falsely assuming screening sensitivity to be constant, study the influence of a higher inter-screening interval and to check the reliability of the estimators under these assumptions.

The paper is organized as follows, Section \ref{s:model} introduces the mathematical part of the model proposed by \cite{Wu}. Section \ref{s:res} shows our estimates of the process parameters based on simulated data. This section is divided into 4 parts, the first corresponds to the results obtained from the simulation based on an exponential sojourn time, in the second we use simulations from a gamma one, in the third from log-logistic sojourn time and in the last one we show the performance of the model when one uses an incorrect distribution to model the sojourn time. In Section \ref{s:prev} we show the reasons behind the discrepancy of estimates in the literature. Finally, the results are discussed in Section \ref{s:diss}.
\section{Model}
\label{s:model}
We will use the general model proposed by   \cite{Wu}. Let us assume we have a cohort of individuals going through a screening program. Suppose that they are stratified by age at program entry and that there are $K$ screens with an inter-screening interval $\Delta$. Let $q(x)$ be the assumed p.d.f. of the sojourn time in the preclinical phase with $Q(t)$ being the survivor function: $Q(t)=\int_t^\infty q(x) dx$ , let $\Phi(t)$ be the sensitivity of the screening exam at age $t$ and define $w(x)$ to be the preclinical intensity.

 Suppose that the first screen $(k=1)$ occurs at $t_0$, define $t_i= t_0 + \Delta \cdot i$  the age of a person at the $i^{th}$  screen, and let $t_{-1}=0$. Then the probability of a person who is aged $t_0$ at the study entry to have cancer detected at the first screen is given by
$$\Phi(t_0) \int_{0}^{{t_0}} w(x) Q(t_0-x)dx$$
And for $k=2, \ldots, K$, the probability of having cancer detected by screening at screen $k$ is given by:
\begin{equation}
\begin{aligned}
     D_{k,t_0} &= \Phi(t_{k-1})\Bigg[ \sum_{i=0}^{k-2}\Big[(1-\Phi(t_i))\cdots (1-\Phi(t_{k-2}))\\
     & \int_{t_{i-1}}^{t_i} w(x) Q(t_{k-1}-x)dx \Big] +\\
    &\int_{t_{k-2}}^{t_{k-1}} w(x) Q(t_{k-1}-x)dx\Bigg]
\end{aligned}
\label{eq:1}
\end{equation}
The first term in the equation corresponds to those who have been falsely screened as negative in the previous screens and stayed in the preclinical state till the $k^{th}$ screen. The second term corresponds to those moving into the preclinical state in the previous screening interval and stayed preclinical till the screen occurs.

The probability of a case becoming incident in the $k^{th}$ screening interval can be derived similarly and is given by: 
\begin{equation}
\begin{aligned}
   I_{k,t_0} = & \Bigg[\sum_{i=0}^{k-2} (1-\Phi(t_i))\cdots (1-\Phi(t_{k-1})) \\
   & \int_{t_{i-1}}^{t_i} w(x)[Q(t_{k-1}-x) - Q(t_{k}-x)]dx\Bigg] \\
  & +\int_{t_{k-1}}^{t_k} w(x)[1-Q(t_k -x)] dx
  \end{aligned}
 \label{eq:2}
\end{equation}
where the first term corresponds to those who have moved to the clinical state before the $ k^{th}$ screening interval and were falsely screened and the second term corresponds to those moving from $S_f$ to $S_c$ during the $ k^{th}$ screening interval.

 Let us assume that in a screening program with $K$ screens for persons within the age range of $[t_{min},t_{max}]$ using the simulated data, a total of $n_{k,t_0}$ persons aged $t_0$ at program entry are screened at screen $k$ and out of them $s_{k,t_0}$ persons were diagnosed by means of screening and $r_{k,t_0}$  persons were becoming incident in the   $k^{th}$ screening interval. Then the likelihood is proportional to: 
\begin{equation}
    L=\prod_{t_0=t_{min}}^{t_{max}} \prod_{k=1}^{K} I_{k,{t_0}}^{r_{k,t_0}} D_{k,{t_0}}^{s_{k,t_0}} (1-D_{k,{t_0}} -I_{k,{t_0}})^{n_{k,t_0}-s_{k,t_0}-r_{k,t_0}}
\end{equation}

To get the maximum likelihood estimates we need to specify parametric models and then they are 
obtained through nonlinear minimization of the negative log-likelihood. The variances of the parameter estimators can be approximated using the observed Fisher information matrix.

\section{Results}
\label{s:res}

Simulations were performed to imitate the results of a breast cancer screening program. For that purpose we simulated the progression of 10000 persons for each age group ($N_{t_0}=10000$) for $t_0 \in$ ($t_{min}=40,T=64$). However, instead of directly simulating screened and symptomatic cases from a binomial distribution with probabilities (\ref{eq:1}) and (\ref{eq:2}), we decided to develop a simulator that assigns an actual sojourn time to each case. As a result, we are able to directly compare the actual sojourn time with the estimated one. For more information about the simulation algorithm see the Appendix.

The distribution for the preclinical intensity is chosen to be log-normal $LN(\mu,s^2)$. The transition probability of breast cancer to the preclinical state was estimated by  \cite{lee}. These were found to be right skewed with a heavy tail, so the log-normal distribution was chosen for having similar properties (\cite{Wu}). The defined values for the parameters are $\mu=3.971$ and $s=2.8$, so these values lead to an average age of transition of around 54 years and a standard deviation of 15 years. 

Note that the preclinical intensity is a sub-density, meaning that it is multiplied by the life time risk. E.g. 1 in 8 women will get breast cancer in the USA although risk factors such as family history significantly raise the risk (\cite{risk}). We will use 15\% in our simulations:
$$w(t)=\frac{0.15}{ts \sqrt{2\pi}} \exp{- \frac{\ln (t-\mu)^2}{2s^2}   }.  $$ 
The choice of a slightly higher lifetime risk is for the sake of simulation. A higher lifetime risk leads to more cases progressing into the preclinical state and therefore increasing the number of screened and clinically detected cases.

When minimizing the --log-likelihood, there is a possibility of getting an unrealistic minimum that has a very large sojourn time, a very low preclinical intensity and a very low sensitivity, so some constraints are definitely needed. For this purpose, the negative log likelihood will be minimized under the constraints $ 3.5 \leq \mu \leq 4.5$ and $0<s \leq 1$. These constraints were derived by assuming that the mode age $ \exp(\mu-s^2)$ of transition into the preclinical state should lie within  $(30, 70) $ as well as assuming that the average age of transition should be less than 70 years: $\exp(\mu + \frac{s^2}{2}) \leq 70$. Combining these two assumptions leads to the bounds on $\mu$ and $ s$ (\cite {Wu}). 

We chose to simulate progression based on 3 different sojourn time distributions, namely :
\begin{enumerate}
    \item An exponential distribution $Exp(\frac{1}{\lambda})$ having a mean sojourn time $\frac{1}{\lambda}=2.5$ years.
    \item A gamma distribution $\Gamma(\alpha, \beta)$ where the shape parameter is $\alpha=6.25$ and a rate parameter $\beta=2.5$ resulting in a mean sojourn time of 2.5 years and unit variance.
    \item A log-logistic $LL(\rho,\kappa)$ with a scale $\rho=4.7$ and shape $\kappa=2.2$ also leading to a mean sojourn time around 2.37 years and approximately a unit variance.
\end{enumerate}
The motivation behind choosing these distributions and their parameters is that we wanted to create very different sojourn time distributions allowing us to see the performance of the model under each chosen distribution.

Breast cancer's screening sensitivity is known to be increasing with age (\cite{shapiro}).  \cite{Wu} choose to model this age dependence via a logistic function with parameters $b_0$ and $b_1$. As a result, the sensitivity at age $t$ is then given by
\begin{equation}
   \Phi(t)=\frac{1}{\exp(-b_0-b_1(t-\bar{t}))}
\end{equation}
where $\bar{t}$ is the average age of entry in the study. The values defined for the simulation were $b_0=1.4$ and $b_1=0.05$ resulting in a sensitivity between 68\% for those aged 40 and 93\% for those aged 64. These values are based on the estimates of \cite{sensitivity}.  Note that $b_1=0$ means that the sensitivity is constant and independent of age. So, in order to see the effects of falsely assuming the sensitivity to be constant, the model will be run twice, first forcing $b_1$ to be 0, and then with no constraints on $b_1$.
 
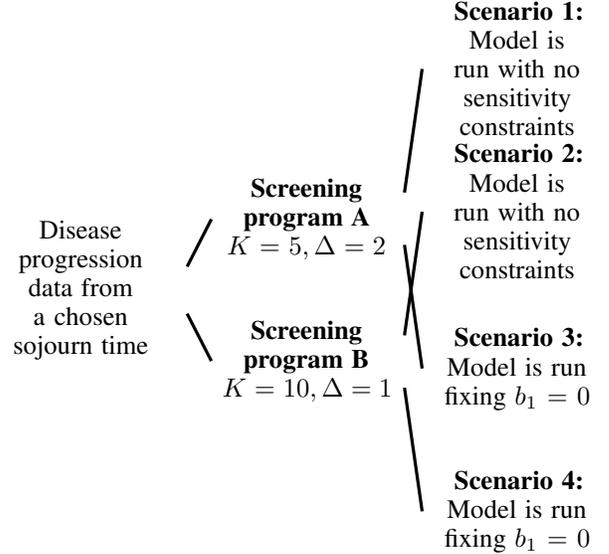
\begin{figure}
\begin{tikzpicture}
  [
    grow=right,
    level 1/.style={sibling distance=1.9cm,level distance=3.cm},
    level 2/.style={sibling distance=4cm, level distance=2.8cm},
    edge from parent/.style={very thick,draw=black,
        shorten >=0.2pt, shorten <=10pt},
    edge from parent path={(\tikzparentnode.east) -- (\tikzchildnode.west)},
    kant/.style={text width=2.3cm, text centered, sloped},
    every node/.style={text ragged, inner sep=1mm},
    punkt/.style={rectangle, rounded corners, shade, top color=white,
    bottom color=black, draw=black, very
    thick }
    ]

  \node[kant]  {Disease progression \\ data from a chosen\\ sojourn time}
    child {node[kant] {\textbf{Screening program B \\ $K=10,\Delta=1$}}
      child {node [kant]{\textbf{Scenario 4:} Model is run fixing $b_1=0$}}
      child {node [kant]{\textbf{Scenario 2:} \\ Model is run with no sensitivity constraints}}
    }
    child {node [kant]{\textbf{Screening program A \\ $K=5,\Delta=2$}}
    child {node [kant]{\textbf{Scenario 3: }Model is run fixing $b_1=0$}}
      child {node[kant] {\textbf{Scenario 1:} \\ Model is run with no sensitivity constraints}}
    };
\end{tikzpicture}
\caption{Scenarios}
\label{fig:scen}
\end{figure}

For each chosen sojourn time distribution, two screening programs were simulated for the same disease progression data, one with $K=5$ screens and an inter-screening interval of $\Delta=2$ years, and the other with $K=10$ ($\Delta=1$ year), hence we can probe the effect of a longer screening interval. The model is run with and without the mentioned sensitivity constraint on each of the data sets, resulting in 4 scenarios shown in Figure \ref{fig:scen}.

The function \textbf {\tt{nlm}} in R is used to minimize the negative of the log-likelihood. The initial values for the minimization are chosen at random, as follows. The initial values for the sensitivity parameters $b_0$ and $b_1$ are chosen from a uniform random variable $U[0,5]$ and $U[0,0.5]$ respectively, these values result in a sensitivity between 0 and 99.99 \% therefore have no effect on the minimization. The preclinical intensity parameters $\mu$ and $s$ are chosen from  $U[3.5]$ and $U[0,1]$ respectively, the choice is based on a the constraints mentioned earlier. The sojourn time parameters are initialized in the following way: $\frac{1}{\lambda}$ from U[0,15] in the exponential case, $\alpha$, $\beta$, $\kappa$ and $\rho$ are all initialized from a $U[0,10]$. The minimization is based on a Newton-Raphson method, it calculates gradient at each step, if the gradient is approximately 0 that means that the current value is probably a solution. The algorithm was run multiple times and it was noticed that the likelihood and the estimators were converging to the same values.

\subsection{Exponentially distributed sojourn time}

\begin{figure*}
\includegraphics[width=\textwidth, height=12cm]{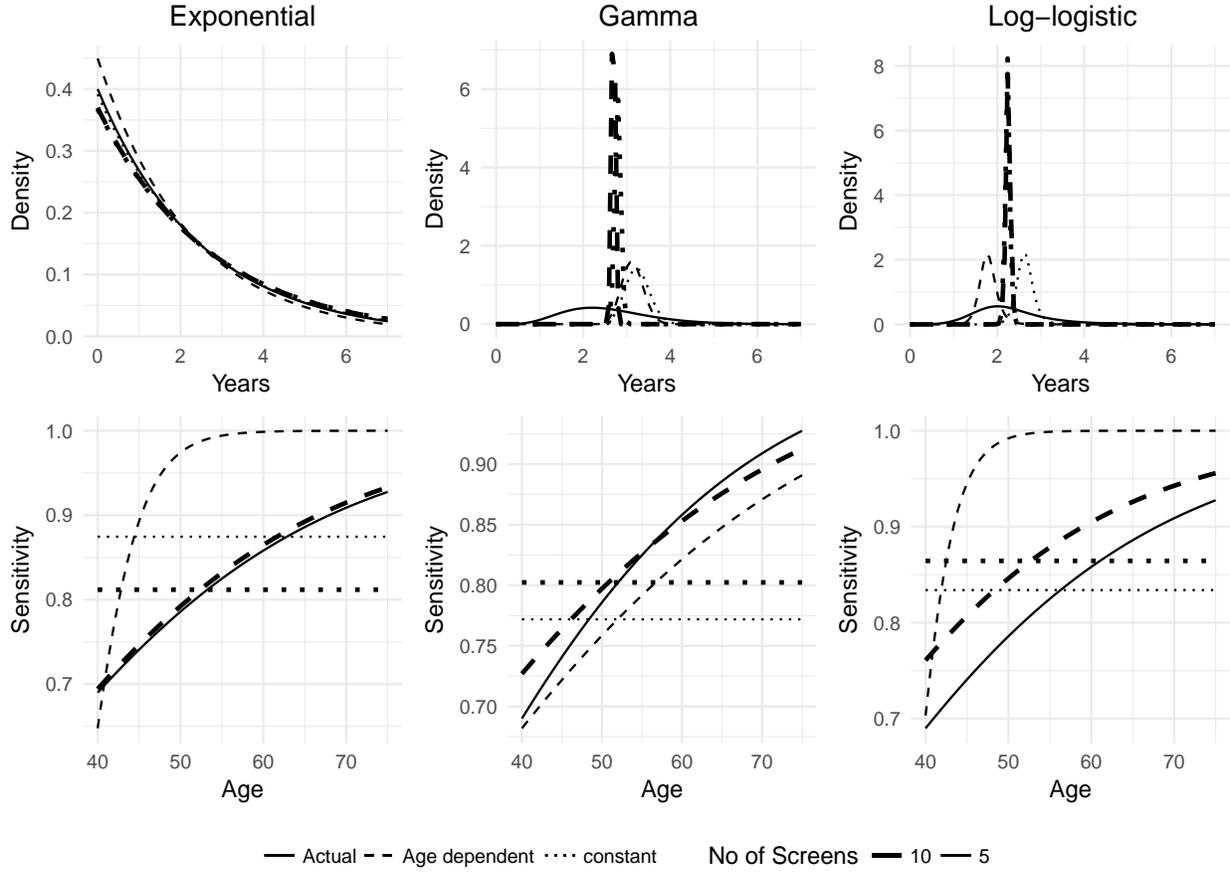}
\caption{Plots for actual and estimated sojourn time density (top) and  sensitivity (bottom) in all scenarios -- data generated from exponential distribution(left), gamma distribution(middle), log-logistic(right)}
\label{fig:all}
\end{figure*}

The exponential distribution is the most commonly used one in the literature. Table \ref{tab:exp} shows the results, the plots for the sojourn time and the sensitivity are shown in the left panels of Figure \ref{fig:all}.

The defined value for the mean sojourn time is $\lambda$= 2.5 years, the one observed from the simulated sample however is around 2.93 years. The structure of the model itself leads to a bias which cannot be captured by the exponential distribution. In fact, on the first screen, the probability of a case to be detected by a screen is determined by those who have moved into the preclinical state before the program started and stayed preclinical till the first screen. In other words, only cases with large enough sojourn time will participate in the first screen, as those who became clinical before that will not participate. This explains the simulated sample mean sojourn time of 2.93 years, since the estimate of the exponential distribution parameter is known to be sensitive to outliers.
\begin{center}
\begin{table*}
\label{tab:exp}
\caption{Estimated sensitivity parameters ($b_0$ and $b_1$),  preclinical intensity parameters ($\mu$ and $s$) and sojourn time parameter ($\lambda$) -- data generated by an exponentially distributed sojourn time}
\begin{tabularx}{\textwidth}{XXXXXX}
\hline
Parameters & $b_0(SD)$		&	$b_1	(SD)$	&	$\mu	(SD)$	&	$s	(SD)$	&	$1/\lambda$	$(SD)$\\
\hline
\textbf{Actual}			&		1.4				&	0.05			&	3.971	&	0.268	&	2.5 	\\

S 1		&		4.238(1.116)	&	0.302(0.109)	&	3.974(0.005)	&	0.269(0.004)	&	2.226(0.057)	\\
S 2		&		1.442(0.144)	&	0.052(0.011)	&	3.972(0.006)	&	0.276(0.004)	&	2.705(0.107)	\\
S 3		&		1.941(0.415)	&	0(NA)			&	3.975(0.006)	&	0.272(0.005)	&	2.554(0.192)	\\
S 4		&		1.462(0.139)	&	0(NA)		&	3.975(0.006)	&	0.273(0.004)	&	2.722(0.107)	\\
\hline
\end{tabularx}
\end{table*}
\end{center}

Results observed from scenario 1 show a very high sensitivity, it can be seen from the bottom part of Figure \ref{fig:all}
that the sensitivity is increasing rapidly with age. The estimated mean sojourn time is 2.2 years, which is less than the actual sojourn time of 2.5 years. The preclinical intensity estimators are exact in all scenarios, there are small fluctuations but they are almost negligible. Moving to scenario 3, which is based on the same run of the screening program, but forcing the sensitivity to be constant, we observe that the estimated sensitivity is around 87\% with a mean sojourn time of 2.554.\\
It can be noticed that the sojourn time and the sensitivity have a strong negative correlation $r=-0.83$. This is contradictory to assuming the sensitivity and the sojourn time to be independent when constructing the model. Screening acts as a censoring mechanism: once a case is detected by a screen, the rest of its sojourn time cannot be observed. This explains scenario 1 results, having high sensitivity with low sojourn time. The same can also be seen in scenario 3 but on a lower scale as the sensitivity is forced to be constant. 

In scenario 2, estimates of the sensitivity parameters are pretty accurate, with a mean sojourn time of around 2.705 years, the estimates are good showing that the model performs well in this scenario. We may compare these values to the results from scenario 4 having an estimated constant sensitivity of around 82\%, and a sojourn time of 2.722 years. It can be noticed that when using constant sensitivity, the estimated one is around the average sensitivity. However, as there is a strong correlation with the sojourn time, choosing to model an age dependent sensitivity using a constant one causes the mean sojourn time to become age dependent. In other words, as younger cases have a lower sensitivity, this creates the false impression that young patients have a higher mean sojourn time, and the older ones have a lower mean sojourn time as they have a higher sensitivity, although the mean sojourn time is the same for both. 
 
Next, we compare the results for different inter-screening intervals. Naturally, a larger inter-screening interval means that there are fewer opportunities for an individual to participate in a screening exam, meaning that it will lead to a high number of clinically detected (interval) cases. The model preserves a good fit in one of two ways, the first is by returning a high sensitivity estimate and a low sojourn time meaning that cases stay a short time in the preclinical state but participation in a screen leads to detection with a high probability. The second is by combining a high sojourn time estimate with a low sensitivity, meaning that cases will stay for a longer time in the preclinical state, therefore having multiple chances to participate in a screen, with screens having a low probability of detection. These can be observed as multiple local minima of the negative log likelihood function.

\subsection{Gamma distributed sojourn time}

Simulating a screening program based on a gamma distributed sojourn time seemed natural due to its relation with the exponential distribution. The results of the scenarios are shown in Table \ref{tab:gamma} with plots for the sojourn time density and sensitivity shown in the middle panels of Figure \ref{fig:all}.

\begin{center}
\begin{table*}
\caption{Estimated sensitivity parameters ($b_0$ and $b_1$),  preclinical intensity parameters ($\mu$ and $s$) and sojourn time parameters ($\alpha$ and $\beta$) -- data generated from a gamma distributed sojourn time}
\label{tab:gamma}
\begin{tabularx}{\textwidth}{XXXXXXXX}
\hline
Scenario	&	$b_0$	&	$b_1$	&	$\mu$	&	$s$	&	$\alpha$	&	$\beta$	&	$\alpha/\beta$	\\
\hline
\textbf{Actual}	&	1.4			&	0.05	&	3.971	&	0.268	&	6.25		&	2.5		&	2.5	\\
S 1&	1.221		&	0.0382	&	3.972	&	0.285	&	156.693		&	50.047	&	3.131	\\
S 2	&	1.447		&	0.039	&	3.971	&	0.276	&	3929.657		&	1458.852	&	2.694	\\
S 3	&	1.219		&	0		&	3.977	&	0.287	&	133.102		&	41.027	&	3.244	\\
S 4	&	1.401		&	0		&	3.975	&	0.277	&	1625.477	&	581.666	&	2.794	\\

\hline
\end{tabularx}

\end{table*}
\end{center}

\begin{figure}
\includegraphics[width=8.5cm ,height=12cm]{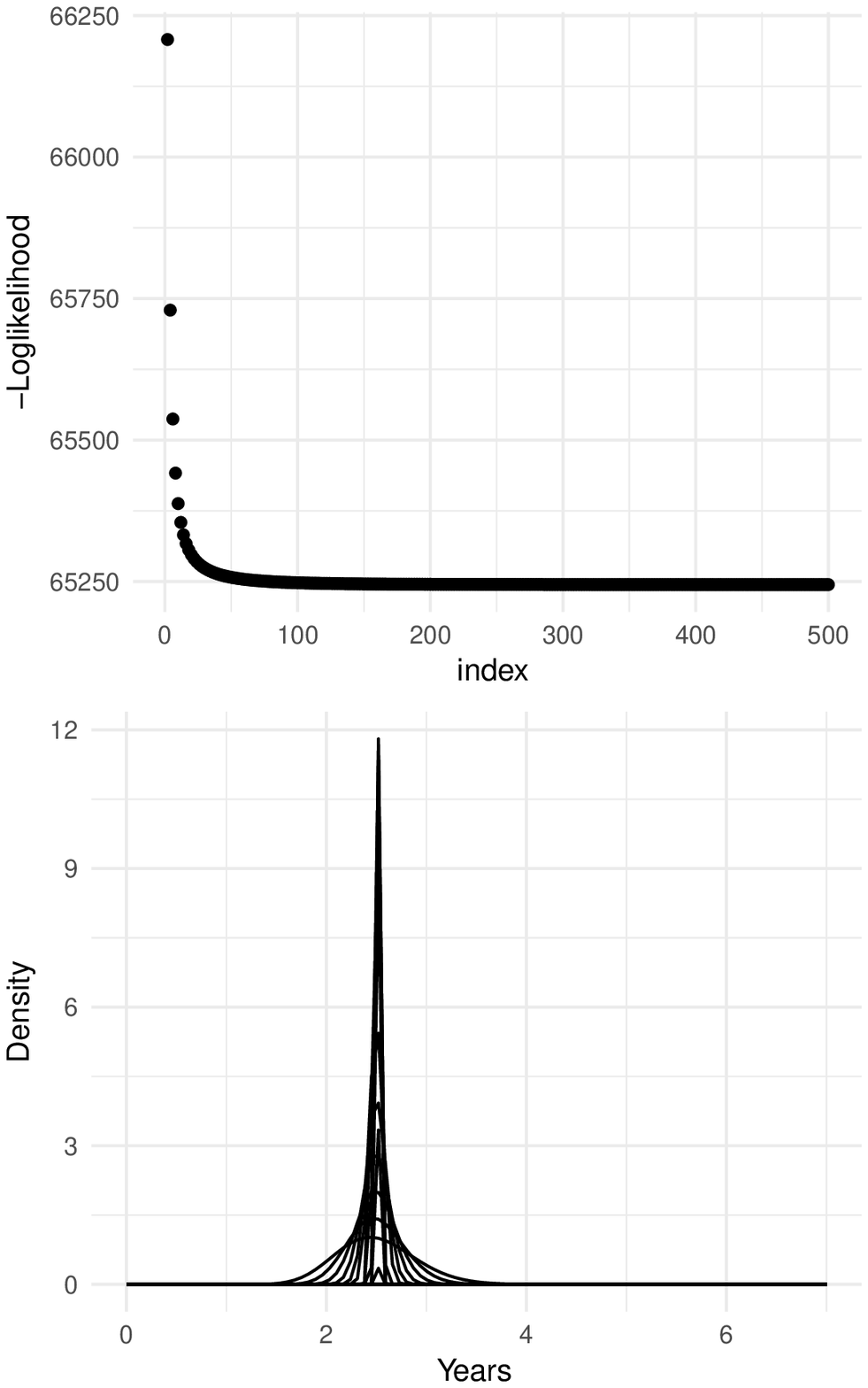}
\caption{Top: -log-likelihood vs index, At the starting point at (index=0) is $\alpha=1\ \& \ \beta=0.4$ and they are updated by adding 1 to $\alpha$ and 0.4 to $\beta$  at each index point (constant mean sojourn  time and decreasing variances), Bottom: equally likely densities}
\label{fig:gamma_lik}
\end{figure}

The first observations are the high values for the sojourn time parameters $\alpha$ and $\beta$ in all scenarios. The upper part of Figure \ref{fig:gamma_lik} shows the negative of the log likelihood while moving on the span of $\alpha =2.5  {\beta}$ starting from $\alpha=1$ and $\beta=0.6$ and updating $\alpha$ by adding 1 and $\beta$ by adding 0.4 maintaining the mean sojourn time $\alpha/ \beta =2.5$ however decreasing the variance $\frac{\alpha}{\beta^2}$. It was noticed that for large $\alpha$ and $\beta$, corresponding to a low variance and a sojourn time which is very dense around its expected value, the likelihood is almost constant and close to the log-likelihood at the maximum. Therefore, these large values of $\alpha$ and $\beta$ will all fall within the acceptable region leading to multiple maxima of the likelihood. This shows a serious flaw in the model. Although it is still able to capture the mean sojourn time under correct assumptions, it is completely unable to capture the variance of the sojourn time, in fact, the model gives equal likelihood for different sojourn time densities  which are very dense around the mean. The bottom part of Figure \ref{fig:gamma_lik} shows multiple densities which have equal likelihood according to the model.

Results also show that the observed Fisher information matrix is not positive definite: it was found that two eigenvalues are very close to 0. Therefore, we are not able to estimate variances of our estimators. This is likely caused by the existence of multiple minima and the strong correlation between the parameters. 
			
That being said, one still gets good estimates for the mean sojourn time and the sensitivity in scenario 2, while forcing the sensitivity to be constant gives slightly worse results in scenario 4. In scenarios 1 and 3, it is noticed that one gets a lower estimate of sensitivity and a higher mean sojourn time. This is due to the larger inter-screening interval in these scenarios.

\subsection {Log-logistic distributed sojourn time}

The log-logistic distribution was suggested by \cite{Wu} as its survivor and hazard functions have relatively simple analytic forms.   E.g. the survivor function of the log-logistic distribution is given by $$Q(x)=\frac{1}{1+(x/\kappa)^\rho}.$$ The results of the scenarios are displayed in Table \ref{tab:loglogis}, plots for the sensitivity and sojourn time are shown in Figure \ref{fig:all}.

The observed results are similar to the gamma case, the observed Fisher information matrix is also not positive definite, although only one eigenvalue is very close to 0. Scenario 2 and 4 seem to perform well, although sensitivity estimates seem to be a bit high. The effect of a high inter-screening interval is amplified here, a higher inter-screening interval either leads to a high sensitivity with a lower sojourn time (as in scenario 1) or a higher sojourn time and a lower sensitivity (as in scenario 2).

\begin{center}
\begin{table*}
\caption{Estimated sensitivity parameters ($b_0$ and $b_1$),  preclinical intensity parameters ($\mu$ and $s$) and sojourn time parameters ($\rho$ and $\kappa$) -- data generated from a log-logistic distributed sojourn time}
\label{tab:loglogis}
\begin{tabularx}{\textwidth}{XXXXXXXX}
\hline
 Scenario	&	$b_0$	&	$b_1$	&	$\mu$	&	$s$	&	$\rho$	&	$\kappa$	&	MST	\\
\hline
\textbf{Actual}	&	1.4		&	0.05	&	3.971	&	0.268	&	2.2		&	4.7		&	2.372	\\
S 1		&	5.623	&	0.396	&	3.969	&	0.256	&	1.785	&	15.506	&	1.797	\\
S 2		&	1.814	&	0.054	&	3.969	&	0.266	&	2.251	&	71.939	&	2.252	\\
S 3		&	1.614	&	0		&	3.974	&	0.274	&	2.633	&	23.187	&	2.641	\\
S 4		&	1.852	&	0		&	3.9713	&	0.264	&	2.238	&	73.774	&	2.239	\\
\hline
\end{tabularx}

\end{table*}
\end{center}

\subsection{Misspecified sojourn time}

One of the main question that this paper answers is what will happen when one chooses an incorrect distribution to model the process. For this purpose, the model is run on the data generated by a known distribution, while intentionally using another distribution to model the sojourn time. In other words, the form of $Q(x)$ is incorrectly chosen to see the effects, this is a realistic situation, as in practice we do not know the data generating process. The models are all run on the data generated by the screening program consisting of 10 screens with $\Delta=1$ with the only constraints being the ones on the preclinical intensity.
\begin{table*}
\caption{Estimated sensitivity parameters ($b_0$ and $b_1$),  preclinical intensity parameters ($\mu$ and $s$) and mean sojourn time (MST) obtained by using different forms of $Q(x)$ (different choices of sojourn time)}
\label{tab:miss2}

\begin{tabularx}{\textwidth}{XXXXXXX}
\hline
\multicolumn{7}{c}{\textbf{Exponential sojourn time }}                                                             \\
$Q(x)$                & Log likelihood & $b_0$         & $b_1$          & $\mu$          & $s$              & MST            \\
\hline
\textit{Actual} & \textit{}      & \textit{1.4}  & \textit{0.05}  & \textit{3.971} & \textit{0.268}   & \textit{2.5}   \\
Exponential     & -69,485.81     & 1.442(0.144)  & 0.052(0.011)   & 3.972(0.006)   & 0.276(0.004)     & 2.705          \\
Gamma           & -69,471.08     & 1.256(0.135)  & 0.033(0.009)   & 3.969(0.005)   & 0.269(0.004)     & 2.541          \\
Log-logistic    & -69,473.60     & 1.316(0.137)  & 0.025(0.008)   & 3.965(0.005)   & 0.267(0.004)     & 2.744          \\\hline

\multicolumn{7}{c}{\textbf{Gamma sojourn time }}       \\
\hline
\textit{Actual} & \textit{}      & \textit{1.4}  & \textit{0.05}  & \textit{3.971} & \textit{0.268}   & \textit{2.5}   \\
Exponential     & -66,191.00     & 1.703 (0.138) & 0.119 (0.013) & 4.028 (0.008)  & 0.330    (0.005) & 4.677 (0.172)  \\
Gamma           & -65,235.42     & 1.447         & 0.039          & 3.971          & 0.276            & 2.694          \\
Log-logistic    & -65,235.42     & 1.447         & 0.039          & 3.971          & 0.276            & 2.694          \\
\hline
\multicolumn{7}{c}{\textbf{Log-logistic sojourn time }}                                                            \\
\hline
\textit{Actual} & \textit{}      & \textit{1.4}  & \textit{0.05}  & \textit{3.971} & \textit{0.268}   & \textit{2.373} \\

Exponential     & -65,928.12     & 2.080 (0.214) & 0.140 (0.021)  & 4.026 (0.007)  & 0.319 (0.005)    & 4.168 (0.161)  \\
Gamma           & -64,781.29     & 1.814         & 0.055          & 3.97           & 0.267            & 2.252          \\
Log-logistic    & -64,781.30     & 1.814         & 0.055          & 3.97           & 0.267            & 2.252          \\ \hline
\end{tabularx}
\end{table*}

Let us start with the data generated from an exponentially distributed sojourn time. The results of fitting different distributions are shown in the top block of Table \ref{tab:miss2}. Using the gamma distribution to model exponentially distributed sojourn time returns $\alpha= 1.269\  (SD=0.0615)$  and $ \beta=0.499\  (SD=0.354)$ giving a mean sojourn time of 2.54, we found that fitting a gamma distribution results in a better fit and more accurate estimates than actually using the exponential one. This is likely due to the ability of the gamma distribution to handle the bias caused by the first screen which was mentioned earlier as well as its ability to capture the exponential distribution (by the parametrization $\alpha=1$). 

Choosing to model an exponential sojourn time with a log-logistic distribution returns a likelihood of -69,473.6, significantly higher than the likelihood of using the exponential distribution. The mean sojourn time estimate is 2.74 years, close to the one estimated by using an exponential distribution. The higher likelihood does not necessarily mean that using the log-logistic distribution is better, as the log-logistic distribution is a two parameter family, it is able to cope with the first screen bias and the exponential one can't therefore the higher likelihood. Yet when comparing the densities, the log-logistic density is very far from the actual density.

Moving on to the data generated from a gamma distributed sojourn time, the results are displayed in the middle block of Table \ref{tab:miss2}. Choosing to model a gamma distributed sojourn time by an exponential distribution to model results in a very high sojourn time estimate of 4.67(0.172) years and it was also noticed that both sensitivity and preclinical intensity parameters are also high with high variances. This is highly problematic as the exponential distribution is the most used one in the literature.

It also seems that using the log-logistic distribution returns a result which is almost identical to the gamma one, with parameters $\rho=2.693$ and $\kappa= 85.455$ both leading to a sojourn time which is extremely dense around the mean sojourn time. 

For data generated from a log-logistic sojourn time, the results are displayed in the bottom block of Table \ref{tab:miss2}. In this case, choosing to model the sojourn time by an exponential distribution results in inaccurate parameter estimates as well, with the mean sojourn time being 4.168 years -- much higher than the real mean sojourn time. Modelling with a gamma distribution returns very large alpha and beta values resulting in a mean sojourn time of 2.252 years. The estimates of preclinical intensity and mean sojourn time are very close to the real ones.

\section{Consequences for previous results}
\label{s:prev}
Table \ref{tab:estimates} shows the estimates of the mean sojourn time and the sensitivity in some famous clinical trials. The health insurance plan of greater New York (HIP) and the clinical trial of Edinburgh \cite{shenzelen2}, the first and the second Canadian National Breast Screening Study (CNBSS1\&2) \cite{Miller1}, \cite{Miller2}.  The CNBSS studies were separated by age, the first for those aged 40--49 at entry and the second for those aged 50--59 at entry, and the Norwegian Breast Cancer Screening Program (NBCSP). One immediately notices the variation between the estimates confirming the sensitivity of the model to the underlying approach, in fact one can notice that there are completely different estimates based on the same data set.

\begin{table*}
\centering
\caption{Sojourn time and sensitivity estimates(M: mammography , P: physical exam ) for some clinical trials}
\label{tab:estimates}
\begin{tabularx}{\textwidth}{lll}
\hline
    	&	Mean sojourn time	&	sensitivity	\\
\hline
HIP (\cite{shenzelen2})	                                    &	2.5	    &	M:0.39 P:0.47
	    \\
Edinburgh (\cite{shenzelen2})	                        	&	4.3	    &	M:0.63, P:0.40    \\
CNBSS1  (\cite{shenzelen2})	                            	&	1.9	    &	M:0.61,  P:0.59	\\
CNBSS2   (\cite{shenzelen2})		                        &   3.1     &	M:0.66,  P:0.39    \\
CNBSS1	 (\cite{Yinlu})         	                        &	2.55	&	0.7	    \\
CNBSS2 (\cite{Yinlu}) 	    	                            &	3.15	&	0.77    \\
NBCSP for the age group [50,59] 	(\cite{lars})	        &	6.1	    &	0.58    \\
NBCSP for the age group [60,69]		(\cite{lars})           &	7.9	    &	0.73    \\
\hline
\end{tabularx}

\end{table*}

 \cite{Yinlu} used a stable disease approach and used the gamma distribution to model the sojourn time of breast cancer. They applied their model on the CNBSS data. They modeled the 40–-49-year-old and 50–-59-year-old cohorts separately, resulting in estimates for the sensitivity for the age groups as 0.70 and 0.77 respectively and for the mean sojourn time as 2.55 years and 3.15 years respectively.  The approach they used assumed the sensitivity and the preclinical intensity  both to be constants. The main issue with their approach is that they chose to bound the shape and the rate parameter of the sojourn time using uniform priors. This may lead to convergence to a local maximum of the likelihood within the chosen bounds. 
The same is done by \cite{Wu}, who used constraints on the sojourn time, the preclinical intensity, as well as the sensitivity when maximizing the likelihood. In other words, they run MCMC simulation on a bounded area to find a local maximum. Although the constraints on the preclinical intensity are natural, constraining the sojourn time has no real justification, especially since the constraints on $\kappa$ and $\rho$ essentially are also constraints on the variance of the sojourn time, something which there is no prior information about. It was shown earlier that the model maintains a constant likelihood when the ratio shape/rate is close to the mean sojourn time.

Now regarding the conflicting results of the CNBSS1 studies, \cite{shenzelen2} estimated the sensitivity for Mammography(M) and physical examination(P) independently, their mean sojourn time estimate for the CNBSS1 trial is 1.9 years, significantly lower than the estimate of \cite{Yinlu} of 2.55 years, they chose to model the sojourn time using a gamma distribution, which is possibly the reason behind the difference in the estimates.

A two parameters (entry--exit) Markov chain model is used by \cite{duffychen}, assuming that the incidence rate $\lambda_1$ and the rate of transition from the preclinical state to the clinical one $\lambda_2$ are both constants. When this method is applied to the data from the Swedish two-county study of breast cancer screening in the age group 70-74, the resulting estimate for the mean sojourn time is 2.3 years. Although the model is very flexible in the sense that interval data is not needed, the constant incidence rate $\lambda_1$ implies that the amount of time spent in the disease free state is an exponential random variable, which is not likely to hold true. In addition to that, the entry--exit model can have multiple feasible maximum likelihood estimates, the extreme case would be a very low preclinical rate combined with a very large sojourn time and a very low sensitivity, applying multiple bounds while maximizing the likelihood is also risky as it may lead to a local maximum within the chosen bounds, possibly excluding the real solution.

Weedon-Fekjaer et al. used a weighted non-linear least-square regression estimates based on a three step Markov chain model, then performed sensitivity analysis to determine the possible impact of opportunistic screening between regular screening rounds. Mean sojourn time and sensitivity were estimated by non-linear least square regression, using number of cancer cases at screening and in the interval between screening examinations. Mean sojourn time was estimated as 6.1 (95\% confidence interval [CI] 5.1-7.0) years for women aged 50-59 years, and 7.9 year (95\% CI 6.0-7.9) years for those aged 60-69 years, sensitivity was estimated as 58\% (95\% CI 52-64 \%) and 73 \% (67-78 \%), respectively. There are multiple issues with these estimates: we suspect that the reason of the high sojourn time estimate is the consequence of the choice of the sojourn time distribution, as we have shown earlier, using the exponential distribution to model the sojourn time if it is not actually exponential results in a very high sojourn time estimate. Their findings also indicate that sensitivity is lower than in  other programs as well as a higher mean sojourn time, we believe it to be a direct consequence of the correlation between the two.

To recreate their results, we ran a simulation for 200,000 initial participants, with the preclinical intensity of 200 per 100000 person years, we simulated a screening program consisting of 5 screens with 2 years inter-screening interval. However, we simulated the sojourn time from a gamma distribution $\gamma(6.25,2.55)$ having a mean sojourn time of 2.5, screens were simulated based on their sensitivity estimate of 58\%.

After generating the data, we modelled the sojourn time using an exponential distribution $(\lambda)$ instead of a gamma one. We then ran the minimization of the -log-likelihood having fixed the sensitivity and the preclinical intensity, in other words, we are only minimizing with respect to the sojourn time parameter $\lambda$. The resulting mean sojourn time estimate was 6.46 years, close to their estimate, but quite far from the real mean sojourn time.

\section{Discussion}
\label{s:diss}

 We can state that the current models are very sensitive to the underlying assumptions. One should take great care when using such an approach and multiple trials with different models are needed before getting results. Summarizing, using the exponential distribution proved to be quite risky as it could result in a much higher mean sojourn time estimates if the data came from another model. It is suggested to use the gamma distribution instead, because it had better properties, not only for the gamma, but also for the exponential model. However multiple minima may exist and it can happen that the model cannot capture the variance of sojourn time. The log-logistic and the gamma distributions behave similarly and models return quite similar results. There is a strong negative correlation between sensitivity and sojourn time, assuming sensitivity to be constant forces the mean sojourn time to be age dependent as a consequence of the correlation between them. Higher inter-screening intervals result in less accurate estimates, either by a high sojourn time and a low sensitivity or the opposite.  
In the future, we plan to model the process incorporating tumor sizes and speed of growth. This may introduce a good constraint on the sojourn time, leading to more stable and accurate results.  

\section*{Acknowledgement}
The project has been supported by the European Union, co-financed by the European Social Fund (EFOP-3.6.2-16-2017-00015).

\section*{Appendix}

\appendix

\section{Simulation Algorithm}
Let us first introduce some notations: denote by $a$ the 99.99\% quantile of the sojourn time, \
 Let $1/h$ be the time discretization constant, $n_{1,t_0}$  the number of individuals who are aged $t_0$ at program entry (first screen). 

Let $t_{0,min}$ and $t_{0,max}$ be the minimum and the maximum age of the target sample at the first screen. Also denote by $V_{t_0}$ the number of preclinical individuals at $t_0$ and by $H_{t_0}=n_{1,t_0}-V_{t_0}$ the number of disease free individuals at $t_0$. Denote by $t$ the individual's age, $\beta(t)$ be the sensitivity of the screening exam at age $t$. Let $w(t)$ be the probability of moving from the disease free state to the preclinical one during $(t,t+dt)$ .

\subsection{Disease progression}
 The first step of the simulation is to determine the initial state of participants, ($H_{t_0}$ and $V_{t_0}$) \\
 for each $t_0 =t_{min}, t_{max}$ and for a predefined $N_{t_0}$.
\begin{enumerate}

\item Assume all individuals are disease free when aged $t_0-b$
\item Discretize the interval  $[t_0-b,t_0)$ into $b*h$ intervals of width $1/h$
\item Denote by $p_{i,h}= \int_{t_0-b+i/h}^{t_0-b+(i+1)/h}w(x)dx$ the probability of moving to the preclinical state in the interval $[t_0-b+i/h,t_0-b+(i+1)/h)$

\item For $i=0,1,\dots,(b*h-1)$ simulate progression into the preclinical state by a $Bernoulli(p_{i,h})$ random variable and store it in an array $x_i$
\begin{itemize} 
	\item If $\sum_0^{b*h-1} x_i=0$ then $H_{t_0}$=$H_{t_0}+1$
	\item If $\sum_0^{b*h-1} x_i\geq 1$ store the age $t_p:=t_0+\min\{i:x_i=1\}*h$ in which the individual became preclinical and generate a random sojourn time $J$ from the chosen sojourn time distribution and assign it.
	
\end{itemize}

\item If $t_p +J \leq t_0$ the case is discarded as it turned out to be clinical before the screening program started.
\item If $t_p +J >t_0$ then  $V_{t_0}$=$V_{t_0}+1$ and store $t_p$, $J$, and $t_0$

\item Stop when $V_{t_0} + H_{t_0}=N_{t_0}$ and repeat the same process for $t_0=t_0+1$
\end{enumerate}

Simulation of progression into the preclinical state during the screening program is done similarly, however checking if $t_p+J<t_0$ is not required and any case becoming preclinical is stored.

\subsection{Screening}

Assume that all individuals go through all $K$ screens. Note that this assumption can be released by simple random sampling out of both healthy and preclinical cases. For a screening program with $K$ screens and an inter-screening time $\Delta$ years the duration of the program is $K*\Delta$ years. 

 For each $t_0 =t_{min},\dots, t_{max}$ and for those who have moved into the preclinical state before $t_0$ define an indicator variable $S$ where $S=1$ indicates detection by screening and $S=0$ indicates clinical detection. Also introduce variable $C=1 ,\dots, K$ which indicates cancer is detected in screen $C$. Note that false positive cases are not of interest as it is assumed that further medical examination will show the absence of the disease. As a result, screening simulation is run only for already preclinical cases. Simulation of screens is done by the following algorithm: start by setting $t=t_0, u=1$
\begin{enumerate} 
\item The detection of cancer by a screen is generated from a $Bernoulli(\beta(t))$
\item If cancer is detected by the screen then set $S=1$ and  $C=u$.
\item If the cancer is not detected by the  screen then  
\begin{itemize}
\item If $t_p+J < t+1$ then the case has become clinical, so store $S=0$ and $C=u$
\item If $t_p+J > t+1$ then update $t=t+1$ and $u=u+1$ and return to step 1
\end{itemize}
\item If the program ends and the case is not detected then the case was neither detected by screening nor by clinical symptoms therefore there was no real information recorded, the case is then just recorded as a person taking a set of screening exams.

\end{enumerate}

For cases developing cancer during the screening program, let us denote $i=1,\dots,K-1$ the interval in which the case became preclinical. One must check if they moved into the preclinical state before the next screen. The simulation is done in a similar way to the stated algorithm, however the number of remaining screens is $K-i$.

\end{document}